\documentclass[12pt,letterpaper]{article}

\usepackage{amsmath}
\usepackage{amsfonts}
\usepackage{amssymb}
\usepackage{graphicx}

\def\be{\begin{eqnarray}}
\def\ee{\end{eqnarray}}

\def\nn{\nonumber}

\newcommand{\di}{\partial}

\begin{document}

\begin{flushright} {\footnotesize IC/2007/032}  \end{flushright}
\vspace{5mm}
\vspace{0.5cm}
\begin{center}

\def\thefootnote{\fnsymbol{footnote}}

{\Large \bf The Effective Field Theory of Inflation}
\\[0.5cm]
{\large Clifford Cheung$^{\rm a}$, Paolo Creminelli$^{\rm b}$,\\[.1cm]
 A.~Liam Fitzpatrick$^{\rm a}$, Jared Kaplan$^{\rm a}$
and Leonardo Senatore$^{\rm a}$}
\\[0.5cm]

{\small
\textit{$^{\rm a}$ Jefferson Physical Laboratory \\
Harvard University, Cambridge, MA 02138, USA}}

\vspace{.2cm}

{\small \textit{$^{\rm b}$ Abdus Salam International Center for
Theoretical Physics\\ Strada Costiera 11, 34014 Trieste, Italy}}

\end{center}

\vspace{.8cm}

\hrule \vspace{0.3cm}
{\small  \noindent \textbf{Abstract} \\[0.3cm]
\noindent
We study the effective field theory of inflation, {\em
  i.e.} the most general theory describing the fluctuations around a
quasi de Sitter background, in the case of single field models. The scalar mode can
be eaten by the metric by going to unitary gauge. In this gauge, the most
general theory is built with the lowest dimension operators invariant
under spatial diffeomorphisms, like $g^{00}$ and $K_{\mu\nu}$, the
extrinsic curvature of constant time surfaces. This approach allows us to
characterize all the possible high energy corrections to simple
slow-roll inflation, whose sizes are constrained by experiments. 
Also, it describes in a common language all single field models, including
those with a small speed of sound and Ghost Inflation, and it 
makes explicit the implications of having a quasi de Sitter background.
The non-linear realization of time diffeomorphisms forces correlation
among different observables, like a reduced speed of sound and an enhanced level of non-Gaussianity.  
\vspace{0.5cm}  \hrule
\def\thefootnote{\arabic{footnote}}
\setcounter{footnote}{0}

\section{Introduction}
The effective field theory approach, {\em i.e.}~the description of a system through the lowest dimension operators compatible with the underlying symmetries, has been very fruitful in many areas, from particle physics to condensed matter. The purpose of this paper is to apply this methodology to describe the theory of fluctuations around an inflating cosmological background.

The usual way to study a single field inflationary model is to start from a Lagrangian for a scalar field $\phi$ and solve the equation of motion for $\phi$ together with the Friedmann equations for the FRW metric. We are interested in an inflating solution, {\em i.e.~}an accelerated expansion with a slowly varying Hubble parameter, with the scalar following an homogeneous time-dependent solution $\phi_0(t)$.
At this point one studies perturbations around this background solution to work out the predictions for the various cosmological observables.  

The theory of perturbations around the time evolving solution is quite different from the theory of $\phi$ we started with: while $\phi$ is a scalar under all diffeomorphisms (diffs), the perturbation $\delta\phi$ is a scalar only under spatial diffs while it transforms non-linearly with respect to time diffs:
\be
t \to t + \xi^0(t, \vec x)   \qquad \delta\phi \to \delta\phi + \dot\phi_0(t) \xi^0 \;.
\ee
In particular one can choose a gauge $\phi(t,\vec x)=\phi_0(t)$ where there are no inflaton perturbations, but all degrees of freedom are in the metric. The scalar variable $\delta\phi$ has been eaten by the graviton, which has now three degrees of freedom: the scalar mode and the two tensor helicities. 
This phenomenon is analogous to what happens in a spontaneously broken
gauge theory. A Goldstone mode, which transforms non-linearly under
the gauge symmetry, can be eaten by the gauge boson (unitary gauge) to
give a massive spin 1 particle. The non-linear sigma model of the
Goldstone can be embedded and UV completed into a linear
representation of the gauge symmetry like in the Higgs sector of the
Standard Model. This is analogous to the standard formulation of
inflation, where we start from a Lagrangian for $\phi$ with a linear
representation of diffs. In this paper we want to stress the
alternative point of view, describing the theory of perturbations
during inflation directly around the time evolving vacuum where time
diffs are non-linearly realized. This formalism has been firstly
introduced, for a generic FRW background, in
\cite{Creminelli:2006xe} to study the possibility of violating the
Null Energy Condition; here we will extend this formalism focusing on
an inflationary solution.

We will show that in unitary gauge the most generic Lagrangian with broken time diffeomorphisms (but unbroken spatial diffs)  describing perturbations around a flat FRW with Hubble rate $H(t)$ is given by
\begin{eqnarray}
\label{eq:Laguni}
S & = & \int  \! d^4 x  \: \sqrt{- g} \;\Big[\frac12 M_{\rm Pl}^2 R+
M_{\rm Pl}^2 \dot{H} g^{00} -  M_{\rm Pl}^2 \left(3
H^2+\dot{H}\right) + \frac{M_2(t)^4}{2!} (g^{00}+1)^2 \\ & & \nonumber + \frac{M_3(t)^4}{3!} (g^{00}+1)^3 + \ldots
-\frac{\bar{M_2}(t)^2}{2}\delta K^\mu {}_\mu {}^2 +... \Big] \; .
\end{eqnarray}
The first two operators after the Einstein-Hilbert term are fixed by
the requirement of having a given unperturbed solution $H(t)$, while
all the others are free and parametrize all the possible different
theories of perturbations with the same background solution. As time diffs are broken one is allowed to write any term that respects spatial diffs, including for example $g^{00}$ and the extrinsic curvature $K^\mu {}_\nu {}$ of the surfaces at constant time. The coefficients of the operators will be in general time dependent.
The reader may be worried by the use of a Lagrangian that is not invariant under diffeomorphisms. But clearly diff. invariance can be restored as in a standard gauge theory. One performs a time-diffeomorphism with parameter $\xi^0(t,\vec x)$ and promotes the parameter to a field $\pi(t, \vec x)$ which shifts under time diffs:
$\pi(t,\vec{x})\rightarrow \pi(t,\vec{x})-\xi^0(t,\vec{x})$. The
scalar $\pi$ is the Goldstone mode which non linearly realizes the time diffs and it describes the scalar perturbations around the FRW solution. 

It is well known that the physics of the longitudinal components of massive gauge bosons can be studied, at sufficiently high energy, concentrating on the scalar Goldstone mode (equivalence theorem). The same is true in our case: for sufficiently high energy the mixing with gravity is irrelevant and we can concentrate on the Goldstone mode. In this regime the physics is very transparent and most of the information about cosmological perturbations can be obtained. Performing the broken diff transformation on the Lagrangian (\ref{eq:Laguni}) and concentrating on the Goldstone mode $\pi$ one gets
\begin{eqnarray} \label{theeffectiveaction} S_\pi = \int \! d^4 x \:
\sqrt{-g}\left[ M_{\rm Pl}^2 \dot H
 \, (\partial_\mu \pi)^2+2 M^4_2\left(\dot\pi^2+ \dot\pi^3-\dot\pi \frac{1}{a^2}(\di_i\pi)^2\right)-\frac{ 4}{3} M^4_3 \dot\pi^3 - \frac{\bar M ^2}{2} \,
\frac{1}{a^4}(\di_i ^2 \pi)^2 +\ldots
\right].
 \end{eqnarray}

Every invariant operator in unitary gauge is promoted to a
(non-linear) operator for the Goldstone: the non-linear realization of
diff invariance forces the relation among various terms.

Let us briefly point out what are the advantages of this approach before moving to a systematic construction of the theory.
\begin{itemize}
\item Starting from a ``vanilla" scenario of inflation with a scalar field with minimal kinetic term and slow-roll potential, we have parameterized our ignorance about all the possible high energy effects in terms of the leading invariant operators. Experiments will put bounds on the various operators, for example with measurements of the non-Gaussianity of perturbations and studying the deviation from the consistency relation for the gravitational wave tilt. In some sense this is similar to what one does in particle physics, where one puts constraints on the size of the operators that describe deviations from the Standard Model and thus encode the effect of new physics.

\item It is explicit what is forced by the symmetries and by the requirement of an inflating background and what is free. For example eq.~(\ref{theeffectiveaction}) shows that the spatial kinetic term $(\nabla \pi)^2$ is proportional to $\dot H$, while the time kinetic term $\dot\pi^2$ is free. Another example is the unitary gauge operator $(g^{00}+1)^2$. Once written in terms of the Goldstone $\pi$, this gives a quadratic term $\dot\pi^2$, which reduces the speed of sound of $\pi$ excitations, and a cubic term $\dot\pi (\nabla\pi)^2$, which increases the interaction among modes, {\em i.e.~}the non-Gaussianity. Therefore, barring cancellations with other operators, a reduced speed of sound is related by symmetry to an enhanced non-Gaussianity. Notice moreover that the coefficient of this operator is constrained to be positive, to avoid propagation of $\pi$ excitations out of the lightcone. 

\item One knows all the possible operators. For example, at the leading order in derivatives, the interaction among three $\pi$ modes can be changed by $(g^{00}+1)^2$ and $(g^{00}+1)^3$. This will correspond to two different shapes of the 3-point function which can be in principle experimentally distinguished to fix the size of each operator. 

\item All the possible single field models are now unified. For
  example there has been interest in models with a modified Lagrangian
  $L((\partial\phi)^2,\phi)$, like DBI inflation
  \cite{Alishahiha:2004eh,Chen:2005ad,Shandera:2006ax,Kecskemeti:2006cg,Shiu:2006kj} which have rather peculiar predictions. In our language these correspond to the case in which the operators $(g^{00}+1)^n$ are large. Another interesting limit is when $\dot H \to 0$; in this case the leading spatial kinetic term is coming from the operator proportional to $\bar M^2$ and it is of the form $(\nabla^2 \pi)^2$. This limit describes Ghost Inflation \cite{Arkani-Hamed:2003uz}.

\item In the $\phi$ language one can perform a field redefinition
  $\phi \to \tilde\phi(\phi)$. It is true that the resulting Lagrangian will
  describe the same physics, but this is not obvious. A simple example
  is given by the Lagrangian 
\be
f(\phi)^2 (\partial\phi)^2 -V(\phi) \;,
\ee
where $f$ is a generic function.
This is equivalent to a Lagrangian with minimal kinetic
term and a different potential through the field redefinition
$\tilde\phi(\phi)$, $d\tilde\phi/d\phi = f(\phi)$. However the
equivalence among different Lagrangians becomes more complicated when
we consider more general terms.
On the other  hand this ambiguity is absent at the level of $\pi$, which realizes
a sort of standard non-linear representation of time diffs.

\item In the $\phi$ language is it not obvious how to assess the importance of an operator for the study of perturbations, because some of the legs of an operator may be evaluated on the background solution. For example in a theory with all operators of the form $(\partial\phi)^{2n}$, all of them may have the same importance if the background velocity $\dot\phi_0$ is large enough, as it happens in DBI inflation.  On the other hand the usual way of estimating the importance of an operator works in the $\pi$ language. Even more clear is the case of Ghost Inflation where, given the non-relativistic dispersion relation for $\pi$ the scaling of operators is clear only in the $\pi$ language.

\item The parametrization of the operators directly around the
  solution is crucial if one calculates loop corrections of
  cosmological perturbations. A diagram with a given number of
  external legs will in general contain a UV divergence. This is easy
  to renormalize in the Lagrangian (\ref{eq:Laguni}), because there is
  only a finite number of terms which describe the interaction among
  $n$ perturbations. On the other hand at the level of the $\phi$
  Lagrangian, there is an {\em infinite} number of operators contributing to
  the interaction among $n$ perturbations. For each operator in fact
  one can put many of its legs on the background, so that the relation
  among an operator and a diagram for perturbations is rather obscure. 
 
\end{itemize}

\section{Construction of the action in unitary gauge}
Inflation is a period of accelerated cosmic expansion with an
approximately constant Hubble parameter. This quasi de Sitter
background has a privileged spatial slicing, given by a physical clock
which allows to smoothly connect to a decelerated hot Big Bang
evolution. The slicing is usually realized by a time evolving scalar
$\phi(t)$. Another example one may keep in mind is given by a perfect fluid \footnote{Indeed, as
shown for example in \cite{Dubovsky:2005xd}, non-vorticous excitations of a perfect fluid may
be described by a derivatively coupled scalar.}. To describe perturbations around this solution one can choose a gauge where the privileged slicing coincides with surfaces of constant $t$, {\em i.e.} $\delta\phi(\vec x,t)=0$. In this gauge there are no explicit scalar perturbations, but only metric fluctuations. As time diffeomorphisms have been fixed and are not a gauge symmetry anymore, the graviton now describes three degrees of freedom: the scalar perturbation has been eaten by the metric. 

What is the most general Lagrangian in this gauge? One must write down
operators that are functions of the metric $g_{\mu\nu}$, and that are invariant
under the (linearly realized) time dependent spatial diffeomorphisms $x^i\rightarrow
x^i+\xi^i(t,\vec{x})$. Spatial diffeomorphisms are in fact
unbroken. Besides the usual terms with the Riemann tensor, which are
invariant under all diffs, many extra terms are now allowed, because
of the reduced symmetry of the system. They describe the additional
degree of freedom eaten by the graviton. For example it is easy to
realize that $g^{00}$ is a scalar under spatial diffs, so that it can
appear freely in the unitary gauge Lagrangian. Polynomials of $g^{00}$
are the only terms without derivatives.  Given that there is a preferred slicing of the spacetime, one is also allowed to write geometric objects describing this slicing. For instance the extrinsic curvature $K_{\mu\nu}$ of surfaces at constant time is a tensor under spatial diffs and it can be used in the action.   Notice that generic functions of time can multiply any term in the action.
In appendix \ref{app:generic} we prove that the most generic Lagrangian can be written as
\begin{eqnarray}
\label{eq:action}\nonumber
S & = & \int  \! d^4 x \; \sqrt{- g} \Big[ \frac12 M_{\rm Pl}^2 R - c(t)
g^{00} -  \Lambda(t) + \frac{1}{2!}M_2(t)^4(g^{00}+1)^2+\frac{1}{3!}M_3(t)^4 (g^{00}+1)^3+ \\
&& - \frac{\bar M_1(t)^3}{2} (g^{00}+1)\delta K^\mu {}_\mu
-\frac{\bar M_2(t)^2}{2} \delta K^\mu {}_\mu {}^2
-\frac{\bar M_3(t)^2}{2} \delta K^\mu {}_\nu \delta K^\nu {}_\mu + ...
\Big] \; ,
\end{eqnarray}
where the dots stand for terms which are of higher order in the fluctuations or with more derivatives.
We denote by $\delta K_{\mu\nu}$ the variation of the extrinsic
curvature of constant time surfaces with respect to the unperturbed
FRW: $\delta K_{\mu\nu} = K_{\mu\nu} - a^2 H h_{\mu\nu}$ with
$h_{\mu\nu}$ is the induced spatial metric. Notice that only the first
three terms in the action above contain linear perturbations around
the chosen FRW solution, all the others are explicitly quadratic or
higher. Therefore the coefficients $c(t)$ and $\Lambda(t)$ will be fixed
by the requirement of having a given FRW evolution $H(t)$, {\em i.e.~}requiring that tadpole terms cancel around this solution. Before fixing these coefficients, it is important to realize that this simplification is not trivial. One would expect that there are an infinite number of operators which give a contribution at first order around the background solution. However one can write the action as a polynomial of linear terms like $\delta K_{\mu\nu}$ and $g^{00}+1$, so that it is evident whether an operator starts at linear, quadratic or higher order. All the linear terms besides the ones in eq.~(\ref{eq:action}) will contain derivatives and they can be integrated by parts to give a combination of the three linear terms we considered plus covariant terms of higher order. This construction is explicitly carried out in appendix \ref{app:tadpoles}.   We conclude that {\em the unperturbed history fixes $c(t)$ and $\Lambda(t)$, while the difference among different models will be encoded into higher order terms.}

We can now fix the linear terms imposing that a given FRW evolution is
a solution. As we discussed, the terms proportional to $c$ and $\Lambda$ are the only
ones that give a stress energy tensor
\begin{equation}
T_{\mu \nu} = -\frac{2}{\sqrt{-g}}\frac{\delta S_{\rm
matter}}{\delta g^{\mu\nu}}
\end{equation}
which does not vanish at
zeroth order in the perturbations and therefore contributes to the
right hand side of the Einstein equations. During inflation we are mostly interested in a flat FRW Universe (see Appendix \ref{app:tadpoles} for the general case) 
\be
ds^2 = -dt^2 + a^2(t) d \vec{x}^2 
\ee
so that Friedmann equations are given by
\begin{eqnarray}
H^2 & = & \frac{1}{3 M_{\rm Pl}^2} \big[ c(t)+\Lambda(t)\big]  \\
\frac{\ddot a}{a} = \dot H + H^2 & =  & -\frac{1}{3 M_{\rm Pl}^2} \big[ 2
c(t)-\Lambda(t) \big] \;.
\end{eqnarray}
Solving for $c$ and $\Lambda$ we can rewrite the action (\ref{eq:action}) as
\begin{eqnarray}
\label{eq:actiontad}\nonumber
S & \!\!\!\!\!\!\!\!\!\!\!\!= \!\!\!\!\!\!\!\!\!& \!\!\!\int  \! d^4 x \; \sqrt{- g} \Big[ \frac12 M_{\rm Pl}^2 R + M_{\rm Pl}^2 \dot H
g^{00} - M_{\rm Pl}^2 (3 H^2 + \dot H) + \frac{1}{2!}M_2(t)^4(g^{00}+1)^2+\frac{1}{3!}M_3(t)^4 (g^{00}+1)^3+ \\
&& - \frac{\bar M_1(t)^3}{2} (g^{00}+1)\delta K^\mu {}_\mu
-\frac{\bar M_2(t)^2}{2} \delta K^\mu {}_\mu {}^2
-\frac{\bar M_3(t)^2}{2} \delta K^\mu {}_\nu \delta K^\nu {}_\mu + ...
\Big] \; .
\end{eqnarray}
As we said all the coefficients of the operators in the action above
may have a generic time dependence. However we are interested in
solutions where $H$ and $\dot H$ do not vary significantly in one
Hubble time.  Therefore it is natural to assume that the same holds
for all the other operators. With this assumption the Lagrangian is
approximately time translation invariant \footnote{The limit in which
  the time shift is an exact symmetry must be taken with care because
  $\dot H \to 0$. This implies that the spatial kinetic term for the
  Goldstone vanishes, as we will see in the discussion of Ghost
  Inflation.}. Therefore the time dependence generated by loop effects
will be suppressed by a small breaking parameter \footnote{Notice that
  this symmetry has nothing to do with the breaking of time
  diffeomorphisms. To see how this symmetry appears in the $\phi$
  language notice that, after a proper field redefinition, one can
  always assume that $\dot\phi =$ const. With this choice, invariance
  under time translation in the unitary gauge Lagrangian is implied by
  the shift symmetry $\phi \to \phi $ + const. This symmetry and the
  time translation symmetry of the $\phi$ Lagrangian are broken down
  to the diagonal subgroup by the background. This residual symmetry
  is the time shift in the unitary gauge Lagrangian.}. This assumption
is particularly convenient since the rapid time dependence of the coefficients can win against the friction created by the exponential expansion, so that inflation may cease to be a dynamical attractor, which is necessary to solve the homogeneity problem of standard FRW cosmology.   

It is important to stress that this approach does describe the
most generic Lagrangian not only for the scalar mode, but also for 
gravity. High energy effects will be encoded for example in operators
containing the perturbations in the Riemann tensor $\delta
R_{\mu\nu\rho\sigma}$. As these corrections are of higher order in
derivatives, we will not explicitly talk about them below.

Let us give some examples of how to write simple models of inflation in this language. A model with minimal kinetic term and a slow-roll potential $V(\phi)$ can be written in unitary gauge as 
\be
\int \! d^4x \: \sqrt{-g}
\left[ -\frac 1 2 (\partial \phi)^2 - V(\phi) \right]  \to  \int \!
d^4x \: \sqrt{- g} \left[ -\frac{\dot \phi_0(t)^2}{2} g^{00} - V(\phi_0(t))
\right] \; .
\ee
As the Friedmann equations give 
$\dot\phi_0(t)^2=-2M^2_P \dot{H}$ and $V(\phi(t))=M_{\rm Pl}^2 (3H^2+\dot H$) we see that the action is of the form (\ref{eq:actiontad}) with all but the first three terms set to zero. Clearly this cannot be true exactly as all the other terms will be generated by loop corrections: they encode all the possible effects of high energy physics on this simple slow-roll model of inflation.

A more general case includes all the possible Lagrangians with at most one derivative acting on each $\phi$: $L= P(X,\phi)$, with $X=g^{\mu\nu}\partial_\mu\phi\partial_\nu\phi$. Around an unperturbed solution $\phi_0(t)$ we have
\be
S = \int \! d^4x \: \sqrt{-g} \;  P(\dot\phi_0(t)^2 g^{00}, \phi(t))
\ee
 which is clearly of the form above with $M_n^4(t) =\dot\phi_0(t)^{2n} \partial^n P/\partial X^n$ evaluated at $\phi_0(t)$.
 Terms containing the extrinsic curvature contain more than one
 derivative acting on a single scalar and will be crucial in the limit
 of exact de Sitter, $\dot H \to 0$.

\section{Action for the Goldstone Boson\label{sec:Goldstone}}
The unitary gauge Lagrangian describes three degrees of freedom: the
two graviton helicities and a scalar mode. This mode will become
explicit after one performs a broken time diffeomorphism
(St\"u{}ckelberg trick) as the Goldstone boson which non-linearly realizes this symmetry. In analogy with the equivalence theorem for the longitudinal components of a massive gauge boson \cite{Cornwall:1974km}, we expect that the physics of the Goldstone decouples from the two graviton helicities at short distance, when the mixing can be neglected.
Let us review briefly what happens in a non-Abelian gauge theory before applying the same method in our case.

The unitary gauge action for a non-Abelian gauge group $A_\mu^a$ is
\be
S = \int \! d^4x  \,-\frac{1}{4} {\rm Tr}\, F_{\mu\nu}F^{\mu\nu}-\frac12 m^2 {\rm Tr}\, A_\mu A^\mu \ ,
\ee
where $A_\mu = A_\mu^a T^a$.
Under a gauge transformation we have
\be
\label{eq:AmuU}
A_\mu \to U A_\mu U^\dagger + \frac{i}{g} U \partial_\mu U^\dagger \equiv \frac{i}{g} U D_\mu U^\dagger \;.
\ee
The action therefore becomes
\be
S = \int \! d^4x  \,-\frac{1}{4} {\rm Tr}\, F_{\mu\nu}F^{\mu\nu} - \frac12 \frac{m^2}{g^2} {\rm Tr} D_\mu U^\dagger D_\mu U \;.
\ee
The gauge invariance can be ``restored" writing $U=\exp{[i T^a \pi^a(t,\vec x)]}$, where $\pi^a$ are scalars (the Goldstones) which transform non-linearly under a gauge transformation $\Lambda$ as
\be
e^{i T^a \widetilde\pi^a(t, \vec x)} = \Lambda(t, \vec x) \,e^{i T^a \pi^a(t,\vec x)}
\ee
Going to canonical normalization $\pi_c \equiv m/g \cdot \pi$, we see that the Goldstone boson self-interactions become strongly coupled at the scale $4 \pi m/g$, which is parametrically higher than the mass of the gauge bosons.
The advantage of reintroducing the Goldstones is that for energies $E \gg m$ the mixing between them and the transverse components of the gauge field becomes irrelevant, so that the two sectors decouple.
Mixing terms in eq.~(\ref{eq:AmuU}) are in fact of the form
\be
\frac{m^2}{g} A_{\mu}^a \partial^\mu \pi^a = m A_{\mu}^a \partial^\mu \pi_c^a
\ee
which are irrelevant with respect to the canonical kinetic term $(\partial \pi_c)^2$ for $E \gg m$. In the window $m \ll E \ll 4 \pi m /g$ the physics of the Goldstone $\pi$ is weakly coupled and it can be studied neglecting the mixing with transverse components.

Let us follow the same steps for our case of broken time diffeomorphisms. Let us concentrate for instance on the two operators:
\begin{equation}
\int d^4x\;  \sqrt{-g} \left[A(t)+B(t)g^{00}(x)\right] \ .
\end{equation}
Under a broken time diff.  $t \to \widetilde t= t + \xi^0(x)$, $\vec{x} \to \vec{\widetilde{x}}=\vec{x}$,  $g^{00}$ transforms as:
\begin{equation}
g^{00}(x)\to \widetilde g^{00}(\widetilde x(x))=\frac{\partial \widetilde x^0(x)}{\partial x^\mu}\frac{\partial \widetilde x^0(x)}{\partial x^\nu} g^{\mu\nu}(x) \,  .
\end{equation}
The action written in terms of the transformed fields is given by:
\begin{eqnarray}
\int d^4x\;  \sqrt{-\widetilde g(\widetilde x(x))} \left|\frac{\partial \widetilde x}{\partial x} \right| \left[A(t)+B(t) \frac{\partial x^0}{\partial \widetilde x^\mu}\frac{\partial x^0}{\partial \widetilde x^\nu} \widetilde g^{\mu\nu}(\widetilde x(x))\right]\ .
\end{eqnarray}
Changing integration variables to $\widetilde x$, we get:
\begin{eqnarray}
\int d^4\widetilde x\;  \sqrt{-\widetilde g(\widetilde x)} \left[A(\widetilde t-\xi^0(x(\widetilde x)))+B(\widetilde t-\xi^0(x(\widetilde x))) \frac{\partial (\widetilde t-\xi^0(x(\widetilde x)))}{\partial \widetilde x^\mu}\frac{\partial (\widetilde t-\xi^0(x(\widetilde x)))}{\partial \widetilde x^\nu} \widetilde g^{\mu\nu}(\widetilde x)\right].
\end{eqnarray}
The procedure to reintroduce the Goldstone is now similar to the gauge theory case. Whenever $\xi^0$ appears in the action above, we make the substitution
\begin{equation} 
\xi^0(x(\widetilde x)) \to - \widetilde \pi(\widetilde x ) \, .
\end{equation}
This gives, dropping the tildes for simplicity:
\begin{eqnarray}
\int d^4x\;  \sqrt{- g(x)} \left[A( t+\pi(x))+B(t+\pi(x)) \frac{\partial (t+\pi(x))}{\partial x^\mu}\frac{\partial (t+\pi(x))}{\partial x^\nu} g^{\mu\nu}(x)\right].
\end{eqnarray}
One can check that the action above is invariant under diffs at all orders (and not only for infinitesimal transformations) upon assigning to $\pi$ the transformation rule
\begin{equation}
\pi(x) \to \widetilde\pi(\widetilde x(x))=\pi(x)-\xi^0(x) \ .
\end{equation}
With this definition $\pi$ transforms as a scalar field plus an additional shift under time diffs.

Applying this procedure to the unitary gauge action (\ref{eq:actiontad}) we obtain
\begin{eqnarray}\label{Smixed}
S = \int \! d^4 x  \: \sqrt{- g} &&\left[\frac{1}{2}M_{\rm Pl}^2 R
- M^2_{\rm Pl} \left(3H^2(t+\pi) +\dot{H}(t+\pi)\right)+ \right.\\ \nn
&&+M^2_{\rm
Pl} \dot{H}(t+\pi)\left(
(1+\dot\pi)^2g^{00}+2(1+\dot\pi)\partial_i\pi g^{0i}+
g^{ij}\partial_i\pi\partial_j\pi\right) + \\ \nonumber
&&\frac{M_2(t+\pi)^4}{2!}\left(
(1+\dot\pi)^2g^{00}+2(1+\dot\pi)\partial_i\pi g^{0i}+
g^{ij}\partial_i\pi\partial_j\pi+1\right)^2 + \nonumber\\
\nonumber && \left. \frac{M_3(t+\pi)^4}{3!}\left(
(1+\dot\pi)^2g^{00}+2(1+\dot\pi)\partial_i\pi g^{0i}+
g^{ij}\partial_i\pi\partial_j\pi+1\right)^3+ ... \right] \; ,
\end{eqnarray}
where for the moment we have neglected for simplicity terms that involve the extrinsic curvature.

This action is rather complicated, and at this point it is not clear what is the advantage of reintroducing the Goldstone $\pi$ from the unitary
gauge Lagrangian. In  analogy with the gauge theory case, the simplification occurs because, at sufficiently short distances, the physics of the Goldstone can be studied neglecting metric fluctuations. As for the gauge theory case, the regime for which this is possible can be estimated just looking at the mixing terms in the Lagrangian above. In eq.(\ref{Smixed}) we see in fact that quadratic terms which mix $\pi$ and $g_{\mu\nu}$ contain fewer derivatives than the kinetic term of $\pi$ so that they can be neglected above some high energy scale. In general the answer will depend on which operators are present.
Let us start with the simplest case in which only the tadpole terms are relevant ($M_2=M_3=\ldots=0$). This corresponds to the standard slow-roll inflation case. The leading mixing
with gravity will come from a term of the form
\begin{equation}
\sim M_{\rm Pl}^2 \dot H \dot\pi \delta g^{00} \ .
\end{equation}


After canonical normalization ($\pi_c\sim M_{\rm Pl} \dot H^{1/2}\pi,\  \delta g_c^{00}\sim M_{\rm Pl} \delta g^{00}$), we see that the mixing terms can be neglected for energies above $E_{\rm mix}\sim \epsilon^{1/2} H$, where $\epsilon$ is the usual slow-roll parameter $\epsilon\equiv -\dot H/H^2$.
Another case which will  be of interest is when the operator $M_2$ gets large. In this case we have mixing terms of the form
\begin{equation}
\sim M_2^4 \dot \pi \delta g^{00}
\end{equation}
which, upon canonical normalization (notice that now $\pi_c\sim M_2^2\pi$), becomes negligible at energies larger than $E_{\rm mix}\sim M_2^2/M_{\rm Pl}$ \footnote{In the theories we are studying Lorentz symmetry is spontaneously broken, so one should define a separate regime of energies and momenta for which the mixing can be neglected. For cosmological perturbations, we will be only interested in the energy range.}.

In the regime $E\gg E_{\rm mix}$ the action dramatically simplifies to
\begin{eqnarray}\label{Spi} 
\! \! S_{\rm \pi} = \int \!  d^4 x   \sqrt{- g} \left[\frac12 M_{\rm Pl}^2 R -M^2_{\rm Pl}
\dot{H} \left(\dot\pi^2-\frac{ (\partial_i \pi)^2}{a^2}\right)
+2 M^4_2
\left(\dot\pi^2+\dot{\pi}^3-\dot\pi\frac{(\partial_i\pi)^2}{a^2}
\right) -\frac{4}{3} M^4_3 \dot{\pi}^3+ ... \right].
\end{eqnarray}

Given an inflationary model, one is interested in computing
predictions for present cosmological observations. From this point of
view, it seems that the decoupling limit (\ref{Spi}) is completely
irrelevant for these extremely infrared scales. However, as for
standard single field slow-roll inflation, one can prove that there
exists a quantity, the usual $\zeta$ variable, 
which is constant out of the horizon at any order in perturbation
theory \cite{Salopek:1990jq,Lyth:2004gb} (see Appendix D of \cite{verification} for a
generalization including terms with higher spatial derivatives). The intuitive reason for the existence of a conserved quantity
is that after exiting the horizon different regions evolve exactly in
the same way. The only difference is how much one has expanded with
respect to another and it is this difference that remains constant.

Therefore the problem is reduced to calculating correlation functions just after horizon crossing. We are therefore interested in studying our Lagrangian with an IR energy cutoff of order $H$.  If the decoupling scale $E_{\rm mix}$ is smaller than $H$, the Lagrangian for $\pi$ (\ref{Spi}) will give the correct predictions up to terms suppressed by $E_{\rm mix}/H$. 

As we discussed, we are assuming that the time dependence of the coefficients in the unitary gauge Lagrangian is slow compared to the Hubble time, that is, suppressed by some generalized slow roll parameters. This implies that  the additional $\pi$ terms coming from the Taylor expansion of the coefficients are small. In particular, the relevant operators, {\it i.e.} the ones which dominate moving towards the infrared, like the cubic term, are unimportant at the scale $H$ and have therefore been neglected in the Lagrangian (\ref{Spi}).

In conclusion, with the Lagrangian (\ref{Spi}) one is able to compute
all the observables which are not dominated by the mixing with
gravity, like for example the non-Gaussianities in standard slow-roll
inflation \cite{Maldacena:2002vr,Seery:2006vu}. Notice however that the tilt of the spectrum can be calculated, at leading order, with the Lagrangian (\ref{Spi}).  As we will see later, its value can in fact be deduced simply by the power spectrum at horizon crossing computed neglecting the mixing terms.
It is important to stress that our approach does not lose its validity
when the mixing with gravity is important so that the Goldstone action
is not sufficient for predictions. The action (\ref{eq:actiontad}) contains all the
information about the model and can be used to calculate all
predictions even when the mixing with gravity is large.

\section{The various limits of single field inflation}

\subsection{Slow-roll inflation and high energy corrections}
The simplest example of the general Lagrangian (\ref{eq:actiontad}) is obtained by keeping only the first three terms, which are fixed once we know the background Hubble parameter $H(t)$, and setting to zero all the other operators of higher order: $M_2 = M_3 = \bar M_1 =\bar M_2 \ldots =0$. In the $\phi$ language, this corresponds to standard slow-roll inflation, with no higher order terms.   
In this case, as discussed in the last section, predictions at the scale $H$ can be made neglecting the mixing with gravity and concentrating on the Goldstone Lagrangian (\ref{Spi}). One is interested in calculating, soon after horizon crossing, the conserved quantity $\zeta$. This is defined, at linear order, by choosing the gauge $\pi = 0$ (unitary gauge in our language) and the spatial part of the metric to be 
\be
\label{eq:zetagauge}
g_{ij} = a^2(t) \left[(1+ 2 \zeta(t,\vec x)) \delta_{ij} + \gamma_{ij}\right]
\ee
where $\gamma$ is transverse and traceless and it describes the two tensor degrees of freedom.
The relation between $\pi$ and $\zeta$ is very simple. As we are neglecting the mixing with gravity, the metric is unperturbed in the $\pi$ language; to set $\pi = 0$ one has to perform a time diffeomorphism
$t \to t - \pi(t,\vec x)$ which gives a spatial metric of the form (\ref{eq:zetagauge}) with
\be
\label{eq:pizeta}
\zeta (t,\vec x) = - H \pi(t, \vec x) \;.
\ee

For each mode $k$, one is only interested in the dynamics around horizon crossing $\omega(k) = k/a \sim H$. During this period the background can be approximated as de Sitter up to slow-roll corrections. Therefore, the 2-point function of the canonically normalized scalar $\pi_c$ is given by the de Sitter result 
\be
\langle \pi_c(\vec k_1) \pi_c(\vec k_2)\rangle = (2 \pi)^3 \delta(\vec k_1 + \vec k_2) \frac{H^2_*}{2 k_1^3} \;,
\ee
 where here and below $*$ means the value of a quantity at horizon crossing. This implies that the 2-point function of $\zeta$ is given by
\be
\label{eq:spectrumzeta}
\langle \zeta(\vec k_1) \zeta(\vec k_2)\rangle = (2 \pi)^3 \delta(\vec k_1 + \vec k_2) \frac{H^4_*}{4 M_{\rm Pl}^2 |\dot H_*|}\frac{1}{k_1^3} = (2 \pi)^3 \delta(\vec k_1 + \vec k_2) \frac{H_*^2}{4 \epsilon_* M_{\rm Pl}^2} \frac{1}{k_1^3} \;.
\ee 
As the variable $\zeta$ is constant outside the horizon, this equation is exact for all $k$ up to slow-roll corrections. In particular  it allows us to calculate the tilt of the spectrum at leading order in slow-roll
\be
n_s-1= \frac{d}{d \log k} \log \frac{H_*^4}{|\dot H_*|} = \frac1{H_*} \frac{d}{d t_*} \log \frac{H_*^4}{|\dot H_*|} = 4 \frac{\dot H_*}{H_*^2} - \frac{\ddot H_*}{H_* \dot H_*} \;.
\ee 

Notice however that not all observables can be calculated from the $\pi$ Lagrangian (\ref{Spi}): this happens when the leading result comes from the mixing with gravity or is of higher order in the slow-roll expansion.  For example, as the first two terms of eq.~(\ref{Spi}) do not contain self-interactions of $\pi$, the 3-point function $\langle \zeta(\vec k_1) \zeta(\vec k_2) \zeta(\vec k_3) \rangle $would be zero. One is therefore forced to  look at subleading corrections, taking into account the mixing with gravity in eq.~(\ref{Smixed}). 

Obviously our choice of setting to zero all the higher order terms
cannot be exactly true. At the very least they will be radiatively
generated even if we put them to zero at tree level. The theory is
non-renormalizable and all interactions will be generated with
divergent coefficients at sufficiently high order in the perturbative
expansion. As additional terms are generated by graviton loops, they
may be very small. For example it is straightforward to check that
starting from the unitary gauge interaction $M_{\rm Pl}^2 \dot H g^{00}$ a
term of the form $(g^{00}+1)^2$ will be generated with a
logarithmically divergent coefficient $M_2^4 \sim \dot H^2 \log
\Lambda$. This implies that one should assume $M^4_2 \gtrsim \dot H^2$
(\footnote{The explicit calculation of logarithmic divergences in a theory
  of a massless scalar coupled to gravity has been carried out a
  long time ago in \cite{'tHooft:1974bx}.}). This lower limit is however very small. For example the dispersion relation of $\pi$ will be changed by the additional contribution to the time kinetic term: this implies, as we will discuss thoroughly below, that the speed of $\pi$ excitations deviates slightly from the speed of light, by a relative amount $1-c_s \sim M_2^4/(|\dot H| M_{\rm Pl}^2) \sim |\dot H|/M_{\rm Pl}^2$. Using the normalization of the scalar spectrum eq.~(\ref{eq:spectrumzeta}), we see that the deviation from the speed of light is $\gtrsim \epsilon^2 \cdot 10^{-10}$. A not very interesting lower limit.

The size of the additional operators will be much larger if additional
physics enters below the Planck scale. In general our approach gives
the correct parametrization of all possible effects of new physics. As
usual in an effective field theory approach, the details of the UV
completion of the model are encoded in the higher dimension
operators. This is very similar to what happens in physics beyond the
Standard Model. At low energy the possible effects of new physics are
encoded in a series of higher dimensional operators compatible with
the symmetries \cite{Barbieri:2004qk}. The detailed experimental study of the Standard model allows us to put severe limits on the size of these higher dimensional operators. The same can be done in our case, although the set of conceivable observations is unfortunately much more limited. 
One example of a possible experimental limit on higher dimension operators is the
consistency relation for the gravitational wave tilt. As is well
known, the gravity wave spectrum from the Einstein-Hilbert action is given by 
\be
\label{eq:GWspectrum}
\langle \gamma^s(\vec k_1) \gamma^{s'}(\vec k_2)\rangle = (2 \pi)^3
\delta(\vec k_1 + \vec k_2) \frac{H_*^2}{M_{\rm Pl}^2}\frac{1}{k_1^3}
\delta_{s s'}\,
\ee 
where $\gamma^s$ denotes the two possible polarizations of the gravity
wave. The ratio between this contribution and the scalar one
(\ref{eq:spectrumzeta}) is given by $\epsilon_*$. The
gravitational wave tilt, $n_g=-2 \epsilon_*$, is thus fixed once the
ratio between tensor and scalar modes is known.

This prediction is valid if one assumes $M_2 =0$, {\em i.e.}
$c_s=1$. As we will see in fact, the scalar spectrum goes as $c_s^{-1}$, while predictions for gravitational waves are not changed by $M_2$. The experimental verification of the consistency relation, even with large errors, would tell us that $c_s$ cannot deviate substantially from $1$ which implies
\be
M_2^4 \lesssim M_{\rm Pl}^2 |\dot H| \;.
\ee

Notice that the higher dimension operators will not only influence
scalar fluctuations, but also the tensor modes, although these
corrections are arguably much harder to test. For example the
unitary gauge operator $-\bar M_3(t)/2  \cdot \delta K^\mu {}_\nu \delta
K^\nu {}_\mu $, whose relevance for scalar fluctuations will be
discussed later on, contains terms of the form $\dot g_{ij}^2$. This
will change the gravity wave dispersion relation. It is in fact
straightforward to obtain the action for the tensor modes
$\gamma_{ij}$ in the presence of this operator. One gets
\be
S_{\gamma}=\frac{M_{\rm Pl}^2}{8} \int d^4x \sqrt{-g} \left[\left(1-\frac{\bar M_3^2}{M_{\rm Pl}^2}\right)\dot\gamma_{ij} \dot\gamma_{ij}- \frac1{a^2}
\partial_l\gamma_{ij} \partial_l\gamma_{ij} \right] \;.
\ee
Therefore the spectrum of gravity waves (\ref{eq:GWspectrum}) will get
corrections of order  $\bar M_3^2/M_{\rm Pl}^2$. This
correction is small unless we push $\bar M_3^2$ up to the Planck
scale. It is easy to realize that operators of the form
$(g^{00}+1)^{n}$ do not influence tensor modes as they do not affect
the transverse-traceless components of the metric.

Other examples of experimental limits on various operators will be discussed in the following sections.

\subsection{Small speed of sound and large non-Gaussianities}
The Goldstone action (\ref{Spi}) shows that the spatial kinetic term $(\partial_i \pi)^2$ is completely fixed by the background evolution to be $M_{\rm Pl}^2 \dot H (\partial_i\pi)^2$. In particular only for $\dot H <0$, it has the ``healthy" negative sign. This is an example of the well studied relationship between violation of the null energy condition, which in a FRW Universe is equivalent to $\dot H<0$, and the presence of instabilities in the system \cite{Hsu:2004vr,Dubovsky:2005xd}. Notice however that the wrong sign of the operator  $(\partial_i \pi)^2$ is not enough to conclude that the system is pathological: higher order terms like $\delta K^\mu {}_\mu {}^2$ may become important in particular regimes, as we will discuss thoroughly below. Reference \cite{Creminelli:2006xe} studies examples in which $\dot H >0$ can be obtained without pathologies.

The coefficient of the time kinetic term $\dot{\pi}^2$ is, on the other hand, not completely fixed by the background evolution, as it receives a contribution also from the quadratic operator $(g^{00}+1)^2$. In eq.~(\ref{Spi}) we have 
\begin{equation}
\left(-M_{\rm Pl}^2 \dot{H} + 2 M_2^4 \right) \dot\pi^2 \;.
\end{equation}
To avoid instabilities we must have $-M_{\rm Pl}^2 \dot{H} + 2 M_2^4 >0$ .
As time and spatial kinetic terms have different coefficients, $\pi$
waves will have a ``speed of sound'' $c_s \neq 1$. This is expected as
the background spontaneously breaks Lorentz invariance, so that
$c_s=1$ is not protected by any symmetry. As we discussed in the last
section, deviation from $c_s=1$ will be induced at the very least by
graviton loops \footnote{If we neglect the coupling with gravity and
the time dependence of the operators in the unitary gauge Lagrangian
(so that $\pi \to \pi + {\rm const}$ is a symmetry),
$c_s=1$ can be protected by a symmetry $\partial_\mu\pi \to
\partial_\mu\pi + v_\mu$, where $v_\mu$ is a constant vector. Under
this symmetry the Lorentz invariant kinetic term of $\pi$ changes by a
total derivative, while the operator proportional to $M_2^4$ in
eq.~(\ref{Spi}) is clearly not invariant, so that $c_s=1$. Notice that
the theory is not free as we are allowed to write interactions with
more derivatives acting on $\pi$. This symmetry appears in the study of the brane
bending mode of the DGP model \cite{Adams:2006sv}.}.
The speed of sound is given by
\be
c_s^{-2} = 1-\frac{2 M_2^4}{M_{\rm Pl}^2 \dot H} \;.
\ee
This implies that in order to avoid superluminal propagation we must
have $M_2^4 >0$ (assuming $\dot H <0$). Superluminal propagation would
imply that the theory has no Lorentz invariant UV completion \cite{Adams:2006sv}. In the following we will concentrate on
the case $c_s \leq 1$, see \cite{Mukhanov:2005bu} for a phenomenological discussion of models
with $c_s > 1$.

Using the equation above for $c_s^2$ the Goldstone action can be
written at cubic order as

\begin{eqnarray}\label{Spi_cs}
S_{\rm \pi} = \int \!  d^4 x  \:  \sqrt{- g} \left[
-\frac{M^2_{\rm Pl} \dot{H}}{c^2_s} \left(\dot\pi^2-c^2_s
\frac{(\partial_i\pi)^2}{a^2}\right)
+M_{\rm Pl}^2 \dot H \left(1-\frac{1}{c^2_s}\right)
\left(\dot{\pi}^3-\dot\pi\frac{(\partial_i\pi)^2}{a^2}
\right)- \frac43 M_3^4 \dot\pi^3... \right].
\end{eqnarray}

From the discussion in section (\ref{sec:Goldstone}) we know that the
mixing with gravity can be neglected at energies $E \gg E_{\rm mix}
\simeq M_2^2/M_{\rm Pl}$. This implies that predictions for cosmological
observables, which are done at energies of order $H$, are captured at leading order by the
Goldstone action (\ref{Spi_cs}) if $H \gg M_2^2/M_{\rm Pl}$, or
equivalently for $\epsilon/c_s^2 \ll 1$. If this is not the case one
is not assured that the Goldstone action contains the leading effects.

The calculation of the 2-point function follows closely the case
$c_s=1$ if we use a rescaled momentum $\bar k=c_s k$ and
take into account the additional factor $c_s^{-2}$ in front of the
time kinetic term. We obtain
\be
\label{eq:spectrumzetacs}
\langle \zeta(\vec k_1) \zeta(\vec k_2)\rangle = (2 \pi)^3 \delta(\vec
k_1 + \vec k_2)  \frac1{c_{s*}} \cdot \frac{H^4_*}{4 M_{\rm Pl}^2 |\dot H_*|}\frac{1}{k_1^3} = (2
\pi)^3 \delta(\vec k_1 + \vec k_2) \frac1{c_{s*}} \cdot \frac{H_*^2}{4 \epsilon_* M_{\rm Pl}^2} \frac{1}{k_1^3} \;.
\ee
The variation with time of the speed of sound introduces an additional
contribution to the tilt
\begin{equation}
n_s=\frac{d}{d \log k}\log \frac{H^4_{*}}{|\dot{H}_*|
c_{s*}}=\frac{1}{H_*}\frac{d}{dt_*}\log \frac{H^4_{*}}{|\dot{H}_*|
c_{s*}} =4\frac{\dot{H_*}}{H^2_*}-\frac{\ddot{H}_*}{\dot{H}_*
H_* }-\frac{\dot{c}_{s*}}{c_{s*}H_{*}} \;.
\end{equation}
The result agrees with the one found in \cite{Garriga:1999vw}.

From the action (\ref{Spi_cs}) we clearly see that the same operator
giving a reduced speed of sound induces cubic couplings of the
Goldstones of the form $\dot{\pi}(\nabla\pi)^2$ and $\dot\pi^3$. The
non-linear realization of time diffeomorphisms forces a relation
between a reduced speed of sound and an enhanced level of
the 3-point function correlator, {\em i.e.} non-Gaussianities. This
relationship was stressed in the explicit  calculation of the
3-point function in \cite{Chen:2006nt}.

To estimate the size of non-Gaussianities, one has to compare
the non-linear corrections with the quadratic terms around freezing, $\omega \sim H$. 
In the limit
$c_s\ll 1$, the operator $\dot\pi(\nabla{\pi})^2$ gives the leading
contribution, as the quadratic action shows that a mode freezes with $k
\sim H/c_s$, so that spatial derivatives are enhanced with respect to
time derivatives. The level of non-Gaussianity will thus be given by the
ratio:
\begin{equation}
\frac{{\cal L}_{\dot\pi(\nabla \pi)^2}}{{\cal L}_{2}}\sim \frac{H
\pi \left(\frac{H}{c_s} \pi\right)^2}{H^2 \pi^2} \sim \frac{H}{c^2_s}
\pi \sim \frac{1}{c^2_s} \zeta \, ,
\end{equation}
where in the last step we have used the linear relationship
between $\pi$ and $\zeta$, eq.~(\ref{eq:pizeta}). Taking $\zeta \sim
10^{-5}$ we have an estimate of the size of the non-linear correction
\footnote{The size of the non-linear corrections depend on the
  specific value of $\zeta$. Even if the typical value of
  $\zeta$ is small, one may be interested in very
large (and therefore very unlikely) fluctuations, for example to study
the production of primordial black holes. For sufficiently large
values of $\zeta$, $\zeta \gtrsim c_S^2$, non-linear corrections become
of order 1 and the perturbative expansion
breaks down. Therefore, predictions which depend on very large values
of $\zeta$ may lie out of the regime of validity of the effective
field theory.}. 
Usually the
magnitude of non-Gaussianities is given in terms of the parameters $f_{\rm
NL}$, which are parametrically of the form: ${\cal L}_{\dot\pi(\nabla
\pi)^2}/{\cal L}_2\sim f_{\rm NL} \zeta$. The leading contribution
will thus give 
\begin{equation}
\label{eq:nabla2dot}
f^{\rm equil.}_{\rm NL, \;\dot\pi(\nabla\pi)^2}\sim \frac{1}{c^2_s} \;.
\end{equation}
The superscript ``equil.'' refers to the momentum dependence of the
3-point function, which in these models is of the so called equilateral form
\cite{Babich:2004gb}. This is physically clear in the Goldstone
language as the relevant $\pi$ interactions contain derivatives, so
that they die out quickly out of the horizon; the correlation is only
among modes with comparable wavelength.

In the Goldstone Lagrangian (\ref{Spi_cs}) there is an
additional independent operator, $-\frac43 M_3^4 \dot\pi^3$,
contributing to the 3-point function, coming from the unitary gauge
operator $(g^{00}+1)^3$. We thus have two contributions of the form
$\dot\pi^3$ which give
\begin{equation}\label{fnldotpicube}
f^{\rm equil.}_{\rm NL,\ \dot\pi^3}\sim 1-\frac{4}{3}
\frac{M_3^4}{M_{\rm Pl}^2 |\dot H| c_s^{-2}} \;.
\end{equation}
The size of the operator $-\frac43 M_3^4 \dot\pi^3$ is not constrained
by the non-linear realization of time diffeomorphisms: it is a free
parameter. In DBI inflation \cite{Alishahiha:2004eh} we have 
$M_3^4 \sim M_{\rm Pl}^2 |\dot H| c_s^{-4}$, so that its contribution to
non-Gaussianities is of the same order as the one of 
eq.~(\ref{eq:nabla2dot}). The same approximate size of the $M_3^4$ is
obtained if we assume that both the unitary gauge operators 
$M_2^4 (g^{00}+1)^2$ and $M_3^4 (g^{00}+1)^3$ become strongly coupled
at the same energy scale.

It is interesting to look at the experimental limits on
non-Gaussianities as a constraint on the size of the unitary gauge operator
$(g^{00}+1)^2$ and therefore on the speed of sound. The explicit
calculation \cite{Chen:2006nt} gives the contribution of the operator $\dot\pi
(\nabla\pi)^2$ to the experimentally constrained parameter $f_{\rm
  NL}^{\rm equil.}$; at leading in order in $c_s^{-1}$ we have
\footnote{This is obtained setting $P_{,XXX}=0$ in the notation of  \cite{Chen:2006nt}.}
\be
f_{\rm NL}^{\rm equil.} = \frac{85}{324} \cdot \frac1{c_s^2} \;.
\ee
The experimentally allowed window \cite{Creminelli:2006rz}
\be
\label{eq:fnleqexp}
-256 < f_{\rm NL}^{\rm equil.} < 332 \quad {\rm at} \; 95\% \;{\rm C.L.}
\ee
translates into the constraint 
\be
c_s > 0.028 \quad {\rm at} \; 95\% \;{\rm C.L.}
\ee
Notice however that, although in principle the operators $\dot\pi (\nabla\pi)^2$
and $\dot\pi^3$ give a different momentum dependence of the 3-point
function, this difference is not experimentally appreciable at present,
so that the constraint (\ref{eq:fnleqexp}) is on the joint effect of the two
operators. The constraint on the speed of sound will hold barring
a cancellation between the two operators. In the case of DBI inflation
for example the effect of the operator $M_3^4 (g^{00}+1)^3$ is
sizeable as we discussed. However there is no cancellation and the
constraint on the speed of sound is only slightly changed to
\be
{\rm DBI:} \quad c_s > 0.031 \quad {\rm at} \; 95\% \;{\rm C.L.}
\ee

Although we concentrated so far on the Goldstone Lagrangian, it is
important to stress that this general approach is useful also when one
is interested in taking into account the full mixing with gravity. For
example, going back to the unitary gauge Lagrangian
(\ref{eq:actiontad}), we can easily see how many coefficients will be
relevant in calculating the 3-point function. At leading order in
slow-roll and in derivatives there are 2 coefficients as we discussed: $M_2$ and $M_3$. At first order in
slow-roll, there will be 4 new parameters describing the slow
variation of the coefficients: the conventional $\epsilon$ and $\eta$
slow-roll parameters and two additional ones for the coefficients of
the 
operators $(g^{00}+1)^2$ and $(g^{00}+1)^3$. This in fact is what one
finds in the explicit calculation \cite{Chen:2006nt} \footnote{The
  explicit calculation shows that one of the coefficients does not
  give rise to an independent momentum dependence of the 3-point
  function, so that it cannot be disentangled from the other parameters.}.

All the discussion can be straightforwardly extended to the 4-point
function (and higher order correlators). In the Goldstone Lagrangian we
have 3 operators contributing to the 4-point function (again at
leading order in slow-roll and derivatives): $(g^{00}+1)^2$,
which is fixed by the speed of sound $c_s$,  $(g^{00}+1)^3$ and
$(g^{00}+1)^4$. Let us estimate the effect of the operator which is
fixed by the speed of sound. As we did for the 3-point function, it is easy to see
that the effect will be dominated by the operator $(\nabla\pi)^4$ and
that the level of non-Gaussianity induced by it can be estimated as
\be
\frac{{\cal L}_{(\nabla \pi)^4}}{{\cal L}_{2}}\sim \frac{\left(\frac{H}{c_s} \pi\right)^4}{H^2 \pi^2} \sim \frac{H^2}{c^4_s}
\pi^2 \sim \frac{1}{c^4_s} \zeta^2 \;.
\ee
This matches with the explicit calculation done in 
\cite{Huang:2006eh}.

\subsubsection{Cutoff and Naturalness}
As discussed, for $c_s <1$ the Goldstone action contains
non-renormalizable interactions. Therefore the self-interactions among
the Goldstones will become strongly coupled at a certain energy scale, which
sets the cutoff of our theory. This cutoff can be estimated looking at
tree level partial wave unitarity, {\em i.e.} finding the maximum energy
at which the tree level scattering of $\pi$s is unitary. The
calculation is straightforward, the only complication coming from the
non-relativistic dispersion relation. The cutoff scale $\Lambda$
turns out to be
\be
\label{eq:cutoff}
\Lambda^4 \simeq 16 \pi^2 M_2^4 \frac{c_s^7}{(1-c_s^2)^2} \simeq 16
\pi^2 M_{\rm Pl}^2 |\dot H|
\frac{c_s^5}{1-c_s^2} \;.
\ee
The same result can be obtained looking at the energy scale where loop
corrections to the $\pi \pi$ scattering amplitude become relevant.
As expected the theory becomes more and more strongly coupled for
small $c_s$, so that the cutoff scale decreases. On the other hand, for
$c_s \to 1$ the cutoff becomes higher and higher. This makes sense as
there are no non-renormalizable interactions in this limit and the
cutoff can be extended up to the Planck scale. This cutoff scale is
obtained just looking at the unitary gauge operator $(g^{00}+1)^2$;
depending on their size the other independent operators may give an
even lower energy cutoff. Notice that the scale $\Lambda$ indicates the maximum energy at which
our theory is weakly coupled and make sense; below this scale new
physics must
come into the game. However new physics can appear even much below
$\Lambda$. 

If we are interested in using our Lagrangian for making predictions for
cosmological correlation functions, then we need to use it at a
scale of order the Hubble parameter $H$ during inflation. We therefore
need that this energy scale is below the cutoff, $H \ll
\Lambda$. Using the explicit expression for the cutoff (\ref{eq:cutoff}) in the case $c_s
\ll 1$ one gets
\be
H^4 \ll M_{\rm Pl}^2 |\dot H| c_s^5
\ee 
which can be rewritten using the spectrum normalization
(\ref{eq:spectrumzetacs}) as an inequality for the speed of sound
\be
c_s \gg P_\zeta^{1/4} \simeq 0.003 \;. 
\ee
A theory with a lower speed of sound is strongly coupled at $E \simeq H$. Not surprisingly this value of the speed of sound also corresponds to the value at which non-Gaussianity are of order one: the theory is strongly coupled at the energy scale $H$ relevant for cosmological predictions.

Let us comment on the naturalness of the theory. One may wonder
whether the limit of small $c_s$ is natural or instead loop
corrections will induce a larger value. The Goldstone
self-interactions, $\dot\pi(\nabla\pi)^2$ and $(\nabla\pi)^4$ for
example, will
induce a radiative contribution to $(\nabla\pi)^2$. It is easy to
estimate that these contributions are of order $c_s^{-5}
\Lambda^4/(16 \pi^2 M_2^4)$, where $\Lambda$ is the UV cutoff, {\em i.e.} the
energy scale at which new physics enters in the game. We can see that
it is impossible to have large radiative contribution; even if we take
$\Lambda$ at the unitarity limit (\ref{eq:cutoff}), the effect is of
the same order as the tree level value. This makes sense as
the unitarity cutoff is indeed the energy scale at which loop
corrections become of order one.

 We would like also to notice that the action (\ref{Spi}) is
natural from an effective field theory point of view
\cite{Polchinski:1992ed}. The relevant operators are in fact protected
from large renormalizations if we assume an approximate shift symmetry
of $\pi$. In this case the coefficients of the relevant operators are sufficiently
small and they will never become important for observations as cosmological correlation functions
probe the theory at a fixed energy scale of order $H$: we never go
to lower energy. Clearly here we are only looking at the period of
inflation, where an approximate shift symmetry is enough to make the
theory technically natural; providing a graceful exit from inflation
and an efficient reheating are additional requirements for a working
model which are not discussed in our formalism.

\subsection{De-Sitter Limit and the Ghost Condensate}

In the previous section we saw that the limit $c_s \to 0$ is
pathological as the theory becomes more and more strongly
coupled. However we have neglected in our discussion the higher derivative
operators in the unitary gauge Lagrangian (\ref{eq:actiontad})
\be
\int d^4 x \sqrt{-g} \left(-\frac{\bar M_2(t)^2}{2} \delta K^\mu {}_\mu {}^2
-\frac{\bar M_3(t)^2}{2} \delta K^\mu {}_\nu \delta K^\nu {}_\mu
\right) \;.
\ee
These operators give rise in the Goldstone action to a spatial kinetic term of the
form
\begin{equation}
\label{di_pi2}  \int  \! d^4x \: \sqrt{-g} \left[ - \frac{\bar M^2}{2} \, \frac{1}{a^4}(\di_i^2 \pi)^2
   \right] \; ,
\end{equation}
where $\bar M^2 = \bar M_2^2+  \bar M_3^2$. 
This spatial kinetic term will make the Goldstone propagate even in
the limit $c_s \to 0$. It is therefore interesting to consider our
general Lagrangian in the limit $\dot H = 0$, when the gravitational
background is exactly de Sitter space which implies $c_s=0$. As $H$ is now time independent,
it is possible to impose an additional symmetry to the theory: the
time independence of all the coefficients in the unitary gauge
Lagrangian. Looking back at the procedure (\ref{Smixed}) to reintroduce the Goldstone
$\pi$, we realize that this symmetry forbids any dependence on $\pi$
without derivatives. The Goldstone action is thus invariant under
shift of $\pi$ 
\be
\pi(\vec x,t) \to \pi(\vec x,t) + {\rm const.}
\ee
 This is the limit of
Ghost Condensation \cite{Arkani-Hamed:2003uy}, where the Goldstone has a non-relativistic
dispersion relation $\omega \propto k^2$. More generally one can
consider intermediate situations where both the spatial kinetic term $c_s^2
(\nabla\pi)^2$ and the higher derivative one $(\nabla^2\pi)^2$ are
present. The predictions of the theory will change significantly
depending on which term dominates at the energy of freezing $\omega
\sim H$ \cite{Arkani-Hamed:2003uz,Senatore:2004rj}.

As with the previous models, one must find the energy regime for which the mixing of the Goldstone with gravity can be neglected.  One simple way to estimate this range is to look at the $\delta K$ operators which contain terms like
\be
\delta K_{ij} \supset (\partial_i\partial_j \pi +\partial_i g_{0j}) \;.
\ee 
Going to canonical normalization this shows that the mixing with gravity can be neglected for 
$k \gtrsim M_2^2/M_{\rm Pl}$. As the dispersion relation of the Goldstone is of the form $\omega^2 = \bar M^2/M_2^4 \cdot k^4$, we see that the energy $E_{\rm mix}$ under which the mixing is relevant is 
$E_{\rm mix} \simeq \bar M M_2^2/M_{\rm Pl}^2$ \cite{Arkani-Hamed:2003uy}.
Notice that this scale has nothing to do with the the curvature of the
background. This is a quite remarkable feature of this example, as
usually the mixing with gravity is related to the background stress
energy tensor and therefore to the curvature of spacetime: the more a
system curves space, the more it mixes with gravity. In this case on
the other hand, the mixing will be relevant even on a flat Minkowski
background. This is what one calls a proper modification of gravity:
gravity, for example the Newtonian potential generated by a source, is
modified at scales much smaller than the curvature. This model of
modification of gravity and its rich phenomenology has been studied in
\cite{Arkani-Hamed:2003uy} \footnote{Also in the case of models with a
  reduced speed of sound, the scale of mixing with gravity can
  become parametrically smaller than the horizon; it is enough to have
  $\epsilon/c_s^2 \ll 1$. In this case the model can be considered a
  way of modifying gravity. Notice however that one can not take the
  limit $\dot H = 0$ without considering the spatial higher derivative
  terms: the scalar mode would not propagate otherwise.}.

As we are interested in inflation, we concentrate on the opposite
limit $H \gg E_{\rm mix}$, when the mixing can be neglected and one
can focus on the $\pi$ Lagrangian. Let us briefly describe the main features of Ghost Inflation, referring for details to 
\cite{Arkani-Hamed:2003uy,Arkani-Hamed:2003uz,Senatore:2004rj}, where
the theory is studied with an approach very close to the one presented
in this paper. Most of the
interesting features can be understood looking at the scaling with
energy of the various operators.  Given the
non-relativistic dispersion relation, $\omega \propto k^2$, the
way an operator scales with energy does not coincide with its mass dimension as in the Lorentz invariant case. 
A rescaling of the energy by a factor $s$, $E\rightarrow s E$,  (equivalent
to a time rescaling $t\rightarrow s^{-1} t$), must go together with a
momentum transformation $k\rightarrow s^{1/2}k$ ($x\rightarrow
s^{-1/2}x$ on the spatial coordinates). As the quadratic action for
the Goldstones is of the form
\be
\label{eq:kinghost}
\int d^4 x \left[ 2 M_2^4 \dot\pi^2 - \frac{\bar M^2}{2} \,
  \frac{1}{a^4}(\di_i^2 \pi)^2 \right]
\ee
we must assign to  $\pi$ the scaling dimension $1/4$
\begin{equation}
\pi\rightarrow s^{1/4}\pi \, .
\end{equation}
to keep the quadratic action invariant. With this rule it is easy to
check that all the allowed Goldstone operators, besides the kinetic
term (\ref{eq:kinghost}), are irrelevant, {\em i.e.} they have
positive scaling dimension and they become less and less relevant
going down in energy. This shows that the theory makes sense as an
effective field theory. In particular the higher derivative
time kinetic operator $\ddot \pi^2$, which would naively seem as
important as $(\nabla^2\pi)^2$ and would describe the presence of a
ghost in the theory, has dimension 2 and it can be neglected
at low energies. If one assumes that there is a single scale $M$ in the
problem, $M \simeq M_2 \simeq \bar M$, this will also set the energy cutoff of the
effective field theory description.

The scaling dimension of $\pi$ also allows us to estimate the spectrum of
perturbations produced in Ghost Inflation. The dimension of $\pi$ tells
us how the amplitude of quantum fluctuations changes with energy. At
the scale of the cutoff $\Lambda \simeq M$, the quantum fluctuations
of the canonically normalized Goldstone field $\pi_c \simeq M^2 \pi $ are of the order of the
cutoff $\delta \pi_c (M)\sim M$. Going down in energy we can estimate
the quantum fluctuations at freezing $E \sim H$. In the standard case
the scalar would have dimension 1 and its fluctuations at freezing
would be of order $H$; in this case on the other hand we have
\begin{equation}
\delta\pi_c(H) \sim \delta\pi_c(M)
\left(\frac{H}{M}\right)^{1/4}\sim \left(H M^3\right)^{1/4} \;.
\end{equation}
Quantum fluctuations at the scale $H$ are much enhanced with respect
to a scalar with a Lorentz invariant dispersion relation. 
The spectrum of $\zeta$ will thus be given by \cite{Arkani-Hamed:2003uz}
\be
\label{eq:spectrumzetaghost}
\langle \zeta(\vec k_1) \zeta(\vec k_2)\rangle \sim (2 \pi)^3 \delta(\vec
k_1 + \vec k_2)  \frac{H^2}{M^4} (H M^3)^{1/2} \frac{1}{k_1^3} \;.
\ee
The correct normalization of the spectrum requires $(H/M)^{5/4} \simeq
10^{-5}$.

The non-Gaussianity will be dominated by the operator with the lowest
dimension. It is straightforward to see that the operator $M_2^4
\dot\pi (\nabla\pi)^2$ coming from the unitary gauge operator $M_2^4
(g^{00}+1)^2$ has dimension $1/4$ and it is the least irrelevant
operator.
At the cutoff scale $M$ the theory is strongly coupled. As the cubic
operator has dimension $1/4$, at energies of order $H$ it will give a
level of non-Gaussianity of order $(H/M)^{1/4}$, which is
parametrically of order $P_\zeta^{1/5}$. The same result can be
obtained with the approach used in the last section, {\em i.e.}
comparing the interaction term with the free action at freezing

\begin{equation}
\label{eq:NGghost}
\frac{{\cal L}_{\dot\pi(\nabla \pi)^2}}{{\cal L}_{2}}\sim
\frac{M^4 H \pi (H M) \pi^2}{M^4 H^2 \pi^2}=M \pi=\frac{M}{H}
\zeta\sim\left(\frac{H}{M}\right)^{1/4} \, .
\end{equation}
The level of non-Gaussianity is extremely high compared to standard
slow-roll as a consequence of the very low dimension of the most
relevant operators. The explicit calculation
\cite{Arkani-Hamed:2003uz} gives an effect which is
somewhat smaller than the naive estimate and comparable to the
existing experimental bound \cite{Creminelli:2006rz}.

In our discussion we have neglected so far the unitary gauge operator
\begin{equation}
\label{eq:leooperator}
\int \! d^4 x \: \sqrt{- g} \:
\left(-\frac{\bar M_1(t)^3}{2} (g^{00}+1)\delta K^\mu {}_\mu \right)\;.
\end{equation}
This operator is odd under time reversal, so that it is consistent to
set it to zero. If this term is present, there is a second
operator with dimension $1/4$ in the Goldstone Lagrangian, of the form $\nabla^2\pi (\nabla\pi)^2$.
Its contribution to the 3-point function would be comparable
with $\dot\pi(\nabla\pi)^2$. The unitary gauge operator
(\ref{eq:leooperator}) also
contributes to the $\pi$ quadratic Lagrangian as we are now going to
discuss.

\subsubsection{De-Sitter Limit without the Ghost Condensate}
In this section we want to study the effect of the operator (\ref{eq:leooperator}) on
the quadratic $\pi$ action. We will see that, if the coefficient of
this operator is sufficiently large, we obtain a new de Sitter limit,
where the dispersion relation at freezing is of the form $\omega^2
\propto k^2$, instead of the Ghost Condensate behavior $\omega^2
\propto k^4$.

For simplicity we can take $\bar M_1$ to be time
independent. Reintroducing the Goldstone we get a 3-derivative term of
the form $-\bar M_1^3 \dot\pi\nabla^2\pi/a^2$ (\footnote{The operator
  gives also a contribution to $\dot\pi^2$ proportional to $H$. We
  will assume that this is small compared to $M_2^4 \dot\pi^2$. In
  Minkowski space the
  operator we are studying can be forbidden by a $\phi \to -\phi$
  symmetry, which is equivalent to time reversal in unitary gauge \cite{Arkani-Hamed:2003uy}. In
  a de Sitter background this symmetry is broken by the metric, so
  that this operator cannot be set to zero.}). This would be a total
time derivative without the time dependence of the scale factor
$a(t)$ and of the metric determinant. Integrating by parts we get a standard 2-derivative spatial
kinetic term
\be
\label{eq:byparts}
-\int d^4 x \sqrt{-g} \,\frac{\bar M_1^3 H}{2}
\left(\frac{\partial_i}{a}\pi\right)^2 \;.
\ee
In the exact de Sitter limit, $\dot H =0$, and taking $M_2 \sim \bar
M_1 \sim M$, this operator gives a dispersion relation of the form
$\omega^2 = c_s^2 k^2$, with a small speed of sound \footnote{In this
  model the mixing with gravity is rather different from the
  previous cases. The reason is that a time derivative is integrated
  by parts to get to eq.~(\ref{eq:byparts}), so that the Goldstone
  terms contain the same number of derivatives as the terms
  describing the mixing with gravity. This implies that the mixing
  does not become less and less relevant going to high energy. On the
  other hand one can choose the model parameters in such a way that
  the mixing is always irrelevant. See \cite{Creminelli:2006rz} for
  the explicit calculations.}

\begin{equation}
c^2_s=\frac{H}{M}\ll 1 \;.
\end{equation}
This will hold only if the higher derivative operators $\delta K^\mu {}_\mu {}^2$ and 
$\delta K^\mu {}_\nu \delta K^\nu {}_\mu$ are subdominant. If we assume
that they are characterized by the same mass scale, $\bar M_2 \sim \bar
M_3 \sim M$, the dispersion relation will get two contributions
\be
\omega^2 \sim \frac{H}{M} k^2 + \frac{k^4}{M^2} \;.
\ee
The two spatial kinetic term are comparable at freezing $\omega \sim
H$. On the other hand, if the $k^4$ contribution is somewhat
suppressed, it becomes irrelevant
 at freezing and therefore for inflationary predictions. In
this limit we have a new kind of Ghost Inflation with an exactly de
Sitter background, but with a $\omega^2 \propto k^2$ dispersion
relationship at freezing. 

Following what we did for the other models it is straightforward to
obtain the spectrum normalization and an estimate of the 3-point
function non-Gaussianity. 
\be
\label{eq:spectrumzetaleo}
\langle \zeta(\vec k_1) \zeta(\vec k_2)\rangle \sim (2 \pi)^3 \delta(\vec
k_1 + \vec k_2)  \frac{H^4}{M^4} \frac1{c_s^3} \frac{1}{k_1^3}  \sim (2 \pi)^3 \delta(\vec
k_1 + \vec k_2)  \left(\frac{H}{M}\right)^{5/2} \frac{1}{k_1^3} \;.
\ee
\be
\frac{{\cal L}_{\dot\pi(\nabla \pi)^2}}{{\cal L}_{2}}\sim
\frac1{c_s^2} H \pi \sim\left(\frac{H}{M}\right)^{1/4} \, .
\ee
A comparable contribution will come from the Goldstone operator
$(\nabla\pi)^2\nabla^2\pi$. Not surprisingly the estimates above are
the same as the ones we obtained in the Ghost Condensate case eqs
(\ref{eq:spectrumzetaghost}) and (\ref{eq:NGghost}). As we discussed
in fact, taking all the operators at the same scale one gets
a comparable contribution at freezing from the $k^2$ and $k^4$ spatial
kinetic terms. We thus expect similar predictions when we assume that
only one of the two contributions is present. 




Now that we have found two different de Sitter limits, one dominated
at freezing by $(g^{00}+1)\delta K^\mu {}_\mu $ and the other by
$\delta K^\mu {}_\mu {}^2$ and $\delta K^\mu {}_\nu \delta K^\nu
{}_\mu$, one may wonder if there are other possibilities. One could
imagine that both these spatial kinetic terms are suppressed for some
reason and the leading operators come at higher order. In this case
one would end up with a dispersion relation of the form
\begin{equation}
\omega^2\sim k^{2n} \quad n \geq 3 \;.
\end{equation}
However it is easy to realize that this cannot be the case, because
the theory would not make sense as an effective field
theory. Following the same logic we used for Ghost Condensation, we
find that the scaling dimension of the operator $\pi$ would be
\begin{equation}
\pi\rightarrow s^{-\frac12+\frac{3}{2n}} \pi \, .
\end{equation}
This implies that the operator $\dot\pi(\nabla\pi)^2$, which is linked
by symmetry to the time kinetic term $\dot\pi^2$, has dimension
$(7-3n)/(2n)$. For $n \geq 3$ this operator is strong at low energy,
so that the effective field theory does not make sense.

\section{Conclusions}
Given the ongoing experimental effort to test inflation and the proliferation of
different models, it is quite important to characterize the most general
theory of inflation. In this paper we took a novel point of view:
instead of writing down a general Lagrangian and study the
fluctuations around an inflating solution, we directly describe the
effective theory of fluctuations around a quasi de Sitter background,
where spatial diffeomorphisms are explicit and the time ones are
non-linearly realized. We showed that the most generic action can be
written at leading order in derivatives in the form
\be
S & \!\!\!\!\!\!\!\!\!\!\!\!= \!\!\!\!\!\!\!\!\!& \!\!\!\int  \! d^4 x \; \sqrt{- g} \Big[ \frac12 M_{\rm Pl}^2 R + M_{\rm Pl}^2 \dot H
g^{00} - M_{\rm Pl}^2 (3 H^2 + \dot H) +
\frac{1}{2!}M_2(t)^4(g^{00}+1)^2+\frac{1}{3!}M_3(t)^4 (g^{00}+1)^3+
\nonumber \\
&& - \frac{\bar M_1(t)^3}{2} (g^{00}+1)\delta K^\mu {}_\mu
-\frac{\bar M_2(t)^2}{2} \delta K^\mu {}_\mu {}^2
-\frac{\bar M_3(t)^2}{2} \delta K^\mu {}_\nu \delta K^\nu {}_\mu + ...
\Big] \; .
\ee

Cosmological correlation functions test this
effective field theory at a scale of order the Hubble parameter
$H$. In this approach the role of symmetries is made much more
transparent. One can see explicitly which features are implied by the
inflating background solution and in particular the
quite different behavior in the cases $\dot H<0$,  $\dot H = 0$ and
$\dot H > 0$ as the coefficient of the operator $g^{00}$ is fixed by
$\dot H$. From this point of view, our approach makes clearer
the relationship among inflation, theories of modification
of gravity and theories which violate the Null Energy Condition
(equivalent to $\dot H >0$ in the cosmological context) like the
bouncing models \cite{Creminelli:2006xe,Buchbinder:2007ad,Creminelli:2007aq}. Another example of the role of symmetries is given by the link between a reduced
speed of sound and an enhanced level of non-Gaussianity as both come
from the same operator $M_2(t)^4(g^{00}+1)^2$ and are thus related by the non-linear
realization of time diffeomorphisms. 

All the possible deviations from a vanilla slow-roll 
scenario are systematically encoded in the size of higher order
operators, similarly to what happens in the study of the Standard
Model of particle physics. Moreover all single field models are
unified in a common framework and this allows us to draw general
conclusions which are independent of the specific realization, as
done in \cite{verification,ArkaniHamed:2007ky} for example.

It is easy to think about possible extensions of our formalism. Along
the same lines it would be interesting to study the most general
theory of (single field) quintessence and to work out its
phenomenological consequences. Differently from inflation, which
probes the effective theory at a scale of order $H$, we would be
interested in this case to the subhorizon dynamics of perturbations.
It would also be interesting to use our approach for the study of
fluctuations in fluids like in radiation or matter dominance \cite{Dubovsky:2005xd}. Finally it should be
straightforward to introduce additional fields into the game and study
multi-field inflationary models.

\section*{Acknowledgments}
It is a pleasure to thank N.~Arkani-Hamed, L.~Boubekeur, S.~Dubovsky,
A.~Guth, F.~Vernizzi, J.~Wacker, M.~Zaldarriaga and especially A.~Nicolis for many useful discussions.
 A. Liam Fitzpatrick is supported by an NSF fellowship. Jared Kaplan is supported by a Hertz fellowship and an NSF fellowship.

\appendix

\section*{Appendix}

\section{The most general Lagrangian in unitary gauge\label{app:generic}}
Let us study what are the rules for writing down the most general Lagrangian in unitary gauge. 
In a theory which is only invariant under spatial diffeomorphisms there is a preferred slicing  of spacetime given by a function $\tilde t(x)$ (with time-like gradient), which non-linearly realizes time diffeomorphisms.
For example if the breaking is given by a time evolving scalar, surfaces of constant $\tilde t$ are also of constant value of the scalar.  Unitary gauge is the one in which the time coordinate $t$ is chosen to coincide with $\tilde t$, so that the additional degree of freedom $\tilde t$ does not explicitly appear in the action. One can therefore build various terms:
\begin{enumerate}
\item Terms which are invariant under all diffeomorphisms: these are just polynomials of the  Riemann tensor $R_{\mu\nu\rho\sigma}$ and of its covariant derivatives, contracted to give a scalar\footnote{The metric and the completely antisymmetric tensor $(-g)^{-1/2} \epsilon^{\mu\nu\rho\sigma}$ can be used to contract indices.}.
\item A generic function of $\tilde t$ becomes $f(t)$ in unitary gauge. We are therefore free to use generic functions of time in front of any terms in the Lagrangian.
\item The gradient $\partial_\mu \tilde t$ becomes $\delta_\mu^0$ in unitary gauge. Thus in every tensor we can always leave free an upper $0$ index. For example we can use $g^{00}$ (and functions of it) in the unitary gauge Lagrangian, or the component of the Ricci tensor $R^{00}$.
\item It is useful to define a unit vector perpendicular to surfaces of constant $\tilde t$
\be
n_\mu = \frac{\partial_\mu\tilde t}{\sqrt{-g^{\mu\nu}\partial_\mu \tilde t \partial_\nu\tilde t}}  \;.
\ee
This allows to define the induced spatial metric on surfaces of constant $\tilde t$: $h_{\mu\nu} \equiv g_{\mu\nu} + n_\mu n_\nu$. Every tensor can be projected on the surfaces using $h_{\mu\nu}$. In particular we can use in our action the Riemann tensor of the induced 3d metric $^{(3)} R_{\alpha\beta\gamma\delta}$ and covariant derivatives with respect to the 3d metric.
\item Additional possibilities will come from the covariant derivatives of $\partial_\mu\tilde t$. Notice that we can equivalently look at covariant derivatives of $n_\mu$: the derivative acting on the normalization factor just gives terms like $\partial_\mu g^{00}$ which are covariant on their own and can be used in the unitary gauge Lagrangian. The covariant derivative of $n_\mu$ projected on the surfaces of constant $\tilde t$ gives the extrinsic curvature of these surfaces
\be
K_{\mu\nu} \equiv h_\mu^\sigma \nabla_\sigma n_\nu \;.
\ee
The index $\nu$ is already projected on the surface because $n^\nu \nabla_\sigma n_\nu = \frac12 \nabla_\sigma (n^\nu n_\nu) =0$. The covariant derivative of $n_\nu $ perpendicular to the surface can be rewritten as 
\be
n^\sigma \nabla_\sigma n_\nu = - \frac12 (-g^{00})^{-1} h^\mu_\nu \partial_\mu(-g^{00})
\ee
so that it does not give rise to new terms. Therefore all covariant derivatives of $n_\mu$ can be written using the extrinsic curvature $K_{\mu\nu}$ (and its covariant derivatives) and derivatives of $g^{00}$.

\item Notice that using at the same time the Riemann tensor of the induced 3d metric and the extrinsic curvature is redundant as $^{(3)} R_{\alpha\beta\gamma\delta}$ can be rewritten using the Gauss-Codazzi relation
as \cite{Wald:1984rg}
\be
^{(3)} R_{\alpha\beta\gamma\delta}= h^\mu_\alpha h^\nu_\beta h^\rho_\gamma h^\sigma_\delta R_{\mu\nu\rho\sigma} -K_{\alpha\gamma}K_{\beta\delta} +K_{\beta\gamma}K_{\alpha\delta} \;.
\ee
Thus one can forget about the 3d Riemann tensor altogether. We can also avoid using the induced metric $h_{\alpha\beta}$ explicitly: written in terms of the 4d metric and $n_\mu$ one gets only terms already discussed above\footnote{Notice that the determinant of the induced metric is related to the one of the full 4d metric by $h = g^{00} \cdot g$ and that the completely antisymmetric 3d tensor can be rewritten in terms of the 4d one as $h^{-1/2} \epsilon^{ijk} = (-g)^{-1/2} (-g^{00})^{-1/2} \epsilon^{0ijk}$.}. Finally also the use of covariant derivatives with respect to the induced 3d metric can be avoided: the 3d covariant derivative of a projected tensor can be obtained as the projection of the 4d covariant derivative \cite{Wald:1984rg}.

\end{enumerate}

We conclude that the most generic action in unitary gauge is given by 
\be
S = \int d^4 x \sqrt{-g} \; F(R_{\mu\nu\rho\sigma}, g^{00}, K_{\mu\nu},\nabla_\mu,t) \;,
\ee
where all the free indices inside the function $F$ must be upper 0's.

\section{Expanding around a given FRW solution\label{app:tadpoles}}
In this Section we want to prove that the most generic theory with broken time diffeomorphisms around a given FRW background (with $k=-1,0,+1$ depending of the spatial curvature) can be written as
\be
S = \int d^4x \sqrt{-g} \left[\frac12 M_{\rm Pl}^2 R + M_{\rm Pl}^2 \Big(\dot H-\frac{k}{a^2}\Big) \cdot g^{00} - M_{\rm Pl}^2 \Big(3 H^2 +\dot H+2\frac{k}{a^2}\Big) + \ldots \right]
\label{eq:general}
\ee
where the dots stand for terms which are invariant under spatial
diffeomorphisms and  of quadratic (or higher) order in the
fluctuations around the given FRW background\footnote{We can always
  make the coefficient in front of $R$ time independent through an
  appropriate field redefinition $g_{\mu\nu} \to g_{\mu\nu} \cdot
  f(t)$. This corresponds, in the usual formalism, to going to Einstein frame.}. 

As such this statement is trivial. We know that the displayed terms give rise to the wanted FRW evolution so that, if we do not want to move away from it, the additional operators must start quadratic around this solution. What we want to say is that {\em each one} of the additional invariant terms is quadratic (or of higher order) without cancellation of linear contributions among various operators. These terms will be written as polynomials (quadratic and higher) of linear operators like $g^{00}+1$, $\delta K_{\mu\nu}= K_{\mu\nu} -K_{\mu\nu}^{(0)}$, $\delta R_{\mu\nu\rho\sigma} \equiv R_{\mu\nu\rho\sigma} - R_{\mu\nu\rho\sigma}^{(0)}$ and so on. Notice that these terms start linear in the perturbations as we have explicitly removed their value evaluated on the given FRW solution. Given the symmetries of a FRW metric, every tensor evaluated on the background ($K_{\mu\nu}^{(0)}$ , $R_{\mu\nu\rho\sigma}^{(0)}$, $(\nabla_\alpha R_{\mu\nu\rho\sigma})^{(0)}$ ...) can be written just in terms of $g_{\mu\nu}$, $n_\mu$ and functions of time. For example 
\begin{eqnarray}
K^{(0)}_{\mu\nu} & =&  a^2 H h_{\mu\nu} \\
R_{\mu\nu\rho\sigma}^{(0)} & = & 2 (H+k) h_{\mu [\rho} h_{\sigma ] \nu} + (\dot H + H^2) a^2 h_{\mu\sigma} \delta_\nu^0 \delta_\rho^0 + {\rm perm.} 
\end{eqnarray} 
where $k$ is a constant which depends on the curvature of the spatial slices and the permutations are acting only on the last term. As such all the operators evaluated on the FRW background are themselves covariant operators, so that operators like $\delta K_{\mu\nu}$ and $\delta R_{\mu\nu\rho\sigma}$ are well defined covariant operators which vanish on the given FRW background and start linear in the perturbations.
We stress that this possibility of rewriting the tensors evaluated on the background holds only because of the high degree of symmetry of the FRW background and it would not be true if one were interested in expanding around a less symmetric solution, {\em e.g.} a non-homogeneous background.
 
In equation (\ref{eq:general}) only the displayed operators contain linear terms in the fluctuations, so that the coefficients of  $\sqrt{-g}\;g^{00}$ and $\sqrt{-g}$ are uniquely determined by the background FRW solution.

Let us now see how the Lagrangian can always be cast in the form (\ref{eq:general}). If we take an operator composed by the contraction of two tensors $T$ and $G$ (the generalization with more tensors is straightforward) we can write
\be
TG = \delta T \delta G + T^{(0)} G + T G^{(0)} - T^{(0)} G^{(0)} \;. \label{eq:splitting}
\ee
 Let us discuss each term of the sum. The first one starts explicitly quadratic in the perturbation as we want. As we said, given the symmetries of the FRW background, the unperturbed tensors $T^{(0)}$ and $G^{(0)}$ can be written as functions of $g_{\mu\nu}$, $n_\mu$ and $t$. Therefore the last term $T^{(0)} G^{(0)}$ is just a polynomial of $g^{00}$ with time dependent coefficients; it contains the terms $\sqrt{-g}\;g^{00}$ and $\sqrt{-g}$ plus operators which start explicitly quadratic in the perturbations.
We are left with tensors of the form $T^{(0)} G$. We want to prove that also these terms can be written as the linear operators in eq.~(\ref{eq:general}) plus operators that start quadratic in the fluctuations. By construction $G$ will be linear either in $K_{\mu\nu}$ or $R_{\mu\nu\rho\sigma}$ with covariant derivatives acting on them.  Covariant derivatives can be dealt with by successive integration by parts, letting them act on $T^{(0)}$ and the time dependent coefficient of the operator. In doing so we can generate extrinsic curvature terms. In this case we can reiterate eq.~(\ref{eq:splitting}) until no covariant derivatives are left\footnote{There can be also powers of $g^{00}$ from $T^{(0)}$. We can deal with them by writing $g^{00} = -1 +\delta g^{00}$ and thus generating additional contributions to the the $g^{00}$ operator in eq.~(\ref{eq:general}) plus terms which are explicitly quadratic or more in the perturbations.}.  We are thus left with the only possible scalar linear terms with no covariant derivatives:  $K^{\mu}  {}_{\mu}$ and $R^{00}$. Both of them can be rewritten in a more useful form. We can integrate by parts the extrinsic curvature term
\begin{eqnarray}
\int d^4x \sqrt{-g}\; f(t) K^{\mu} {}_{\mu} =\int d^4x \sqrt{-g}\; f\; \nabla_{\mu} n^{\mu}=-\int d^4 x \sqrt{-g} \;n^\mu \partial_\mu f =\int d^4x \sqrt{-g} \sqrt{-g^{00}} \dot f \,. 
\end{eqnarray}
While we can deal with $R^{00}$ using the following relationship \cite{Wald:1984rg}:
\begin{eqnarray}
(-g^{00})^{-1} R^{00}=R_{\mu\nu}n^\mu n^\nu = K^2 -K_{\mu\nu}K^{\mu\nu}-\nabla_\mu(n^\mu\nabla_\nu n^\nu)+\nabla_\nu(n^\mu \nabla_\mu n^\nu)\, .
\end{eqnarray}
The last two terms can again be integrated by parts:
\begin{eqnarray}
\int d^4x \sqrt{-g} f(t) \nabla_\mu(n^\mu\nabla_\nu n^\nu)=-\int d^4x \sqrt{-g}\; \partial_\mu f\; n^\mu K^\nu {}_\nu\, ,
\end{eqnarray}
\begin{eqnarray}
\int d^4 x \sqrt{-g} f(t) \nabla_\nu(n^\mu \nabla_\mu n^\nu)=-\int d^4 x \sqrt{-g}\; \partial_\nu f\; n^\mu\nabla_\mu n^\nu =0
\end{eqnarray}
where in the last passage we have used that $\partial_\nu f \propto n_\nu$. This shows that $K^\mu_\mu$ and $R^{00}$ can be written in terms of the linear operators of eq.~(\ref{eq:general}) plus invariant terms that
starts quadratically in the fluctuations.

In conclusion, we have shown that the most general Lagrangian of a theory with broken time diffeomorphisms around a given FRW  background can be written in the form:
\begin{eqnarray}
S &=& \int d^4x \sqrt{-g} \; \Big[\frac12 M_{\rm Pl}^2 R + M_{\rm Pl}^2 \Big(\dot H-\frac{k}{a^2}\Big) \cdot g^{00} - M_{\rm Pl}^2 \Big(3 H^2 +\dot H+2\frac{k}{a^2}\Big) +  \\ \nonumber &&  F^{(2)}(g^{00}+1,\delta K_{\mu\nu}, \delta R_{\mu\nu\rho\sigma};\nabla_\mu;t)\Big]
\label{eq:general2}
\end{eqnarray}
where $F^{(2)}$ starts quadratic in the arguments $g^{00}+1$, $\delta K_{\mu\nu}$ and $\delta R_{\mu\nu\rho\sigma}$.

\end{document}